# Nonergodicity and central limit behavior for systems with long-range interactions

Alessandro Pluchino and Andrea Rapisarda[*]

Dipartimento di Fisica e Astronomia and Infn sezione di Catania, Università di Catania,
Via. S. Sofia 64, 95123, Catania, Italy

## ABSTRACT

In this paper we discuss the nonergodic behavior for a class of long-standing quasi-stationary states in a paradigmatic model of long-range interacting systems, i.e. the HMF model. We show that ensemble averages and time averages for velocities probability density functions (pdfs) do not coincide and in particular the latter exhibit a tendency to converge towards a q-Gaussian attractor instead of the usual Gaussian one predicted by the Central Limit Theorem, when ergodicity applies.

## 1. INTRODUCTION

It is a common practice in statistical physics to exchange time averages with ensemble averages since it is usually assumed that the ergodic hypothesis is in general valid. Although the latter is very often verified it is not always true, expecially for complex systems. In this paper we briefly discuss one example where this happens. We present new numerical simulations for the Hamiltonian mean field (HMF) model [1-5] and we show that, in the out-of-equilibrium regime for long-standing quasi-stationary states characterized by anomalous dynamics, the probability density functions (pdfs) of the two ensembles are different. In particular time averages have an attractor, whose shape is very well reproduced by a q-Gaussian, a pdf typical of generalized statistics [6,7,4]. A more complete discussion on this topic can be found in Refs. [10,11].

## 2. THE MODEL

The HMF model is a paradigmatic model for long-range interacting systems whose dynamics is highly nontrivial. It describes a system of $N$ fully-coupled classical inertial XY spins (rotators) $\vec{s}_i = \left(\cos(\vartheta_i), \sin(\vartheta_i)\right)$, $i = 1, 2, ..., N$, with unitary module and mass [1]. These spins can also be thought as particles rotating on the unit circle. The Hamiltonian is given by

$$H = \sum_{i=1}^{N} \frac{p_i^2}{2} + \frac{1}{2N} \sum_{i,j=1}^{N} 1 - \cos(\vartheta_i - \vartheta_j) \quad , \quad (1)$$

where $\vartheta_i$ is the angle and $p_i$ the conjugate variable representing the rotational velocity of spin $i$. The exact equilibrium solution of the model predicts a second order phase transition when the system passes from a high temperature paramagnetic phase to a low temperature ferromagnetic one [1] increasing the energy density $U=E/N$. The critical temperature is $T_c=1/2$ and corresponds to a critical energy per particle $U_c = E_c/N = 3/4$. The order parameter of this phase transition is the modulus of the *average magnetization* per spin defined as: $M = (1/N) | \sum_i \vec{s}_i |$. The critical exponents are those expected for a mean field model [1]. Above $T_c$, the spins point towards different directions so that $M \sim 0$. Below $T_c$, most spins are aligned, the rotators being trapped into a single cluster, and $M > 0$. The out-of-equilibrium dynamics of the model is also very interesting. In particular in a range of energy densities $U \in [0.5, 0.75]$ and adopting special initial conditions called *water bags*, with initial magnetization $M_0=1$ (i.e. with all the spins aligned and with all the available energy in the kinetic form), the system, after a violent relaxation, goes towards metastable quasi-stationary states (QSS). The latter slowly decay towards equilibrium with a lifetime which diverges like a power of the system size $N$ [2-4]. In the QSS regime many dynamical anomalies have been found as for example vanishing Lyapunov exponents, anomalous diffusion, aging and glassy behavior [2-4]. In the following we presents new numerical results obtained exploring the out-of-equilibrium regime of the HMF model which confirm recently published results [10].

*andrea.rapisarda@ct.infn.it

## 3. DISCUSION OF NUMERICAL RESULTS

In this section we discuss the dynamical evolution of the HMF model for a very large size and an energy density where QSS exist. We construct probability density functions of quantities expressed as a finite sum of stochastic variables. But in this case, we select these variables along the deterministics time evolutions of the N rotators [10]. More formally, we study the pdf of the quantities

$$y_j = \frac{1}{\sqrt{n}} \sum_{i=1}^{n} \left( p_j(i) - \langle p \rangle \right) \quad for \quad j = 1, 2, ... N, \qquad (2)$$

where $p_j(i)$, with $i=1,2,...,n$, are the velocities of the $jth$-rotator, taken at fixed intervals of time $\delta$ along the same trajectory, obtained integrating the HMF equations of motions. See Ref. [3] for details about the integration algorithm adopted. The quantity $<p> = (1/Nn) \sum_j^N \sum_i^n p_j(i)$ is the time average of the $p_j(i)$'s over all rotators and trajectories. Please notice that in Ref. [10] there is a sum over N missing in the average velocity <p> of eq.(2). The product $\delta \times n$ gives the total simulation time over which the sum is calculated. Note that the variables $y$'s are proportional to the *time average* of the velocities along the single rotators trajectories. In the following we distinguish this kind of average from the standard *ensemble average* of the velocities calculated for the N rotators at a given time and over many different realizations of the dynamics. The latter can also be obtained considering the $y$'s variables with $n=1$ and $<p>=0$.

In general, although the standard CLT predicts a Gaussian shape for sum of $n$ independent stochastic values strictly when $n \to \infty$, in practice a finite sum converges quite soon to the Gaussian shape and this, in absence of correlations, is certainly true at least for the central part of the distribution. We discuss in this section a sum of $n=10000$ values of velocities along the deterministic trajectories. We have already shown, in Ref. [10], that if correlations among velocities are strong enough and the system is in weakly chaotic regime, then CLT predictions are not verified and, consistently with recent generalizations of the CLT, $q$-Gaussians appear [8, 9, 11]. The latter are a generalization of Gaussians which emerge in the context of the formalism of nonextensive statistical mechanics proposed by Tsallis [6-7] and are defined as

$$G_q = A \left[ 1 - (1-q) \beta x^2 \right]^{\frac{1}{1-q}} . \qquad (3)$$

In this expression $q$ is the so-called *entropic index* (for $q = 1$ one recovers the usual Gaussian), $\beta$ another suitable parameter which characterizes the width of the distribution and $A$ a normalization constant. One can see also ref. [15] for a simple and general way to generate them. In Ref. [10] we discussed results for systems up to N=50000, here we present new numerical simulations for a system larger than those discussed previously, i.e. for $N=100000$. This fact will allow us to extend the sum along the deterministic trajectories since the lifetime of the QSS is larger. Going to large sizes is in general a very hard task since the cpu time required easily exceeds several weeks. We could actually perform such calculations using new Grid facilities recently available in Catania. For all the present simulations, water-bag initial conditions with initial magnetizazion $M_0 = 1$, usually referred as M1 initial conditions, were used. The energy density adopted is as usual $U = 0.69$, this is in fact the value for which anomalous dynamics is most evident. In fig. 1 we plot the temperature of the system, calculated as twice the kinetic energy per particle, as a function of time. It is evident the initial violent relaxation and the following plateau well below the predicted equilibrium temperature, indicated as a dashed line. The relaxation regime towards this equilibrium value is not shown in the figure. In figs.2 and 3 we plot the pdfs of the quantities $y$ defined in (2).

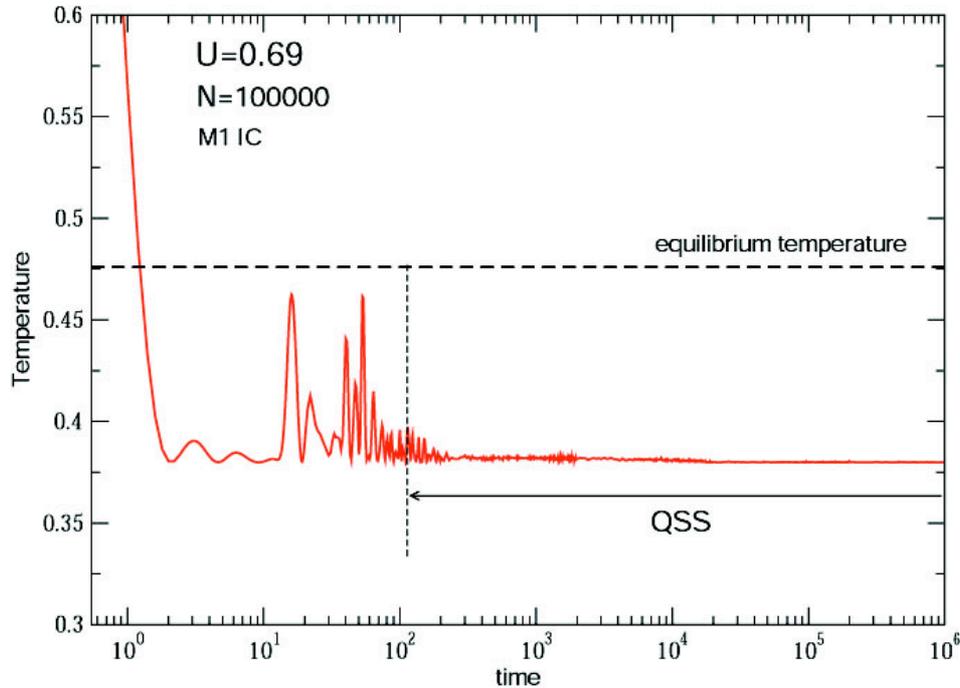

Fig.1 We plot the temperature, calculated as twice the kinetic enery divided by N vs time for the case N=10000 and U=0.69.

The inequivalence between ensemble average and time average is evident comparing these two figures . In fact, in fig.2 we plot the *ensemble average pdf* of the velocities calculated (over 10 different realizations) at time $t = 1000$, i.e. in the QSS regime (full circles), see also fig.1. In this case a peculiar non-Gaussian shape with a central peak and two lateral bumps emerges and remains stable along the QSS regime. A Gaussian curve with unitary variance is also plotted for comparison as dot-dashed curve. Fig.2 must be compared with fig.3, where we plot the *time average pdf* for the normalized variable $y$ with $n = 10000$ and $\delta = 100$, after a transient of 200 time units and over a simulation time of 1000000 units along the QSS. It is important to stress that in this case *only one single realization* of the initial conditions was performed. The shape of the time average pdf results to be very different from that of the ensemble average. In fact it presents very large and robust tails which can be reproduced by a a $q$-Gaussian, defined in (3), with the entropic index $q = 1.4 \pm 0.05$. The different results obtained also for this large size and for a larger values of n confirms the inequivalence between the two kind of averages and the existence of a $q$-Gaussian attractor in the QSS regime of the HMF model. Please notice that in order to allow a correct comparison with standard Gaussians (represented as dot-dashed lines in all the figures) and $q$-Gaussians (represented as full lines), the pdf curves were always normalized to unit area and unit variance, by subtracting from the $y$'s their average $< y >$ and dividing by the correspondent standard deviation $\sigma$ (hence, the traditional $\sqrt{n}$ scaling adopted in Eq. (2) is in fact irrelevant).

Finally, in fig.4, we plot the same curves of fig.3 but this time in linear scale in order to show that also the central part of the pdf and not only the tails are well reproduced by the q-Gaussian curve.

The discrepancy between time averages and ensemble averages is essentially due to a nonergodic behavior in the QSS regime. In fact as illustrated in fig.5, for the case N=10000 and U=0.69, the initial violent relaxation induces a non homogeneous filling of phase space. Many clusters are formed and their size distribution presents a very interesting hierarchical structure. These clusters are in competition among each other in capturing more and more particles, so that a sort of dynamical frustration is generated which inhibits relaxation towards equilibrium, see Ref. [4-5] and references therein for more details.

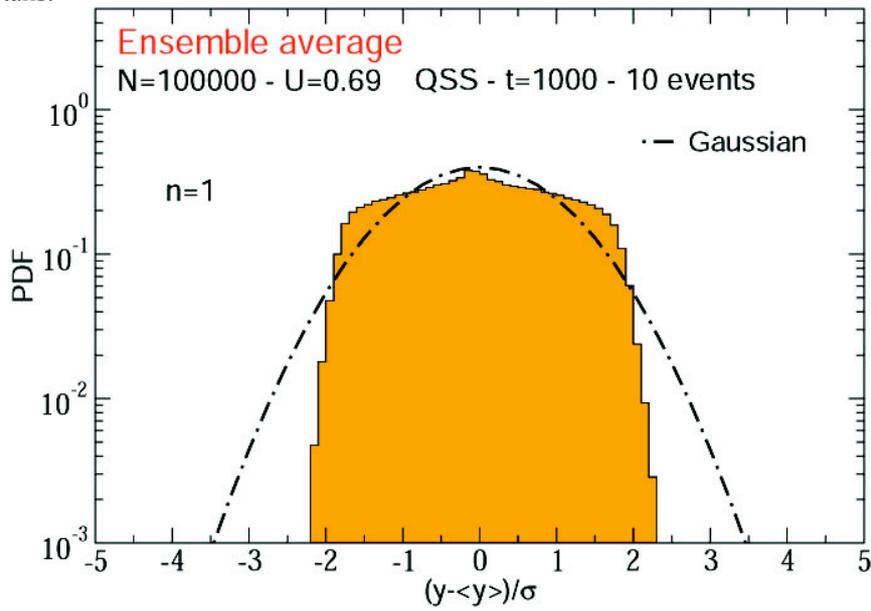

Fig. 2. Ensemble average for N=100000 and U=0.69. The Gaussian pdf (dashed curve) is also plotted for comparison, see text.

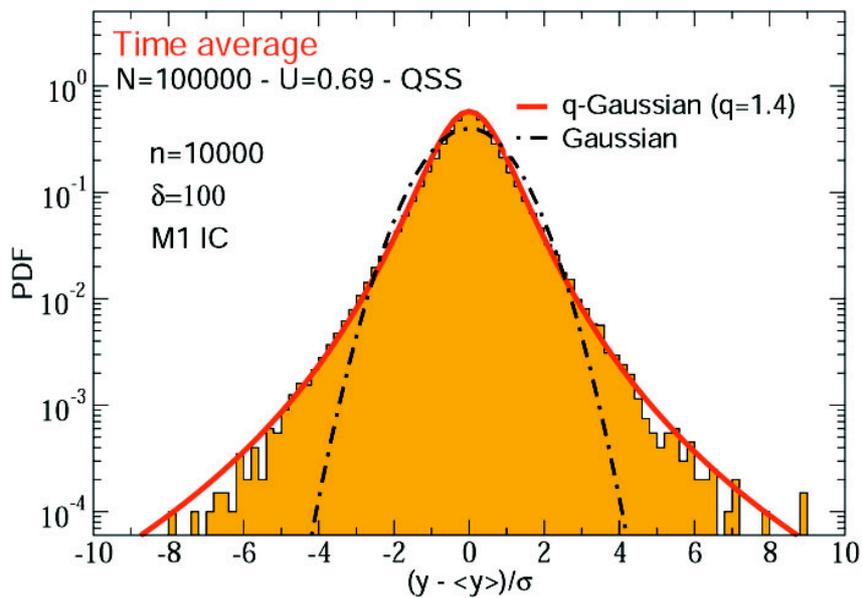

Fig.3. Time averages for for N=10000 and U=0.69, considering n=10000 and δ=100. A Gaussian pdf (dot-dashed curve) and a q-Gaussian with q=1.4 (full curve) are plotted for comparison, see text.

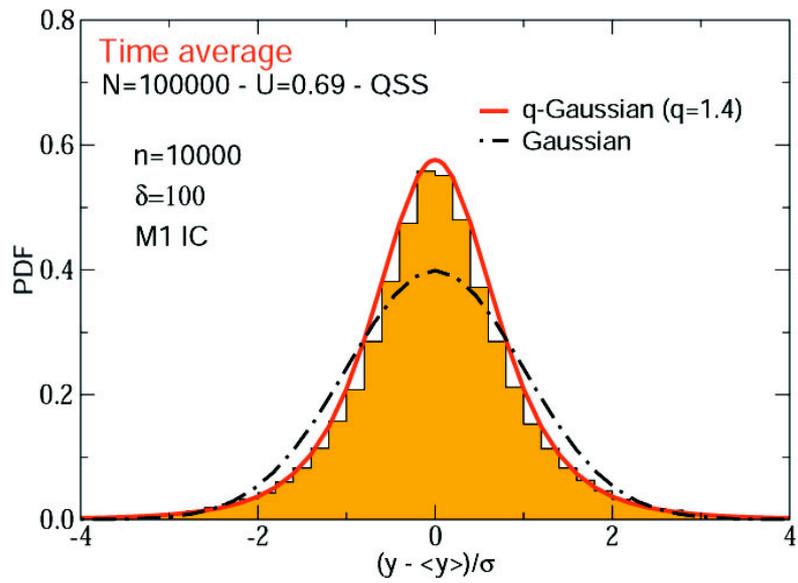

Fig. 4. The central part of the previous figure 3 but htis time plotted using a linear scale. Again the red curve is a q-Gaussian with q=1.4, while the dot-dashed one is a standard Gaussian curve.

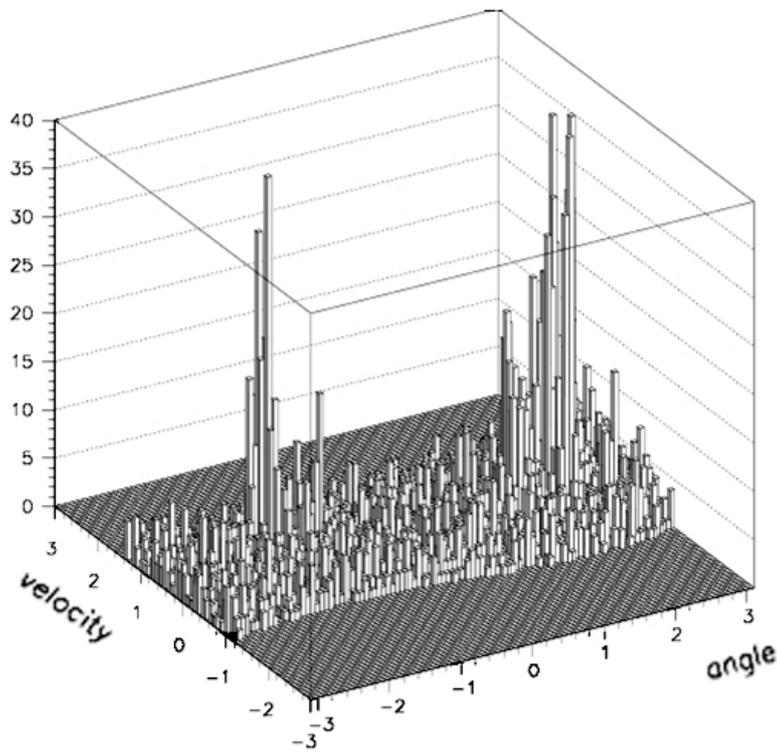

Fig. 5. For the case U=0.69, N=10000 and M1 initial conditions, a 3-dimensional plot of the μ-space occupancy in the QSS regime illustrating the nonergodic behavior due to the presence of clusters with a hierarchical structure. See text.

As a final remark we notice that in a recent paper [16] we have observed three different classes of events in the QSS regime of the HMF model for each given fixed size N. It is then interesting to see in future studies how this different behavior could influence the results here discussed.

## 4. CONCLUSIONS

The numerical simulations presented in this paper confirm for a larger system size recent results [10] and strongly indicate that dynamical correlations and ergodicity breaking, induced in the HMF model by the initial out-of equilibrium violent relaxation, are present along the entire QSS metastable regime and decay very slowly even after it. In particular, considering finite sums of $n$ correlated variables (velocities in this case) selected with a constant time interval $\delta$ along single trajectories, allowed us to study this phenomenon in a very clear and stringent way. Indeed, we have shown that, in the weakly chaotic QSS regime, (i) ensemble average and time average of velocities are inequivalent, hence the ergodic hypothesis is violated, (ii) the standard CLT is violated, and (iii) robust $q$-Gaussian attractors emerge, although it is certainly nontrivial to prove analytically whether the attractor in the nonergodic QSS regime of the HMF model precisely is a $q$-Gaussian or not. In this respect, analytical results, as well as numerical dangers, have been recently presented in Ref.[13] for various models. On the contrary, when no QSS exist, or at a very large time after equilibration, i.e., when the system is fully chaotic and ergodicity has been restored, the ensemble average of velocities results to be equivalent to the time average and one observes a convergence towards the standard Gaussian attractor, see Ref.[10]. In this case, the predictions of CLT are satisfied, even if we have only considered a finite sum of stochastic variables. How fast this happens depends on the size $N$, on the number $n$ of terms summed in the $y$ variables and on the time interval $\delta$ considered. These results are consistent with the recent $q$-generalized forms of the CLT discussed in the literature [8, 9,11,12,14], and pose severe questions to the often adopted procedure of using ensemble averages instead of time averages. Along similar lines, nonergodicity was recently exhibited in shear flows, with results that were fitted with Lorentzians, i.e., $q$-Gaussians with $q = 2$ [17]. The whole scenario reminds that found for the leptokurtic returns pdf in financial markets [18], or in turbulence [19], among many other complex systems.

## ACKNOWLEDGEMENTS

We would like to thank Marcello Iacono Manno for technical support in the preparation of the scripts to run our codes on the GRID platform. The numerical calculations here presented were done within the TRIGRID project. We acknowledge financial support from the PRIN05-MIUR project "Dynamics and Thermodynamics of Systems with Long-Range Interactions". This work is part of a project in collaboration with C. Tsallis.